# Copolymer-induced stabilizing effect of highly swollen hexagonal mesophases

*Laurence Ramos\* and Christian Ligoure*

Laboratoire des Colloïdes, Verres et Nanomatériaux (UMR CNRS-UM2 n°5587), CC26, Université Montpellier II, 34095, Montpellier Cedex 5, France

ramos@lcvn.univ-montp2.fr

Corresponding author: ramos@lcvn.univ-montp2.fr; Phone: +33 467144284; Fax: +33 467144637

ABSTRACT: We show that small amounts of copolymer that decorates a oil/water interface can greatly enhance the stability of swollen surfactant hexagonal phases, comprising oil tubes regularly arranged in a water matrix. Both the radius of the tubes and the thickness of the aqueous channel between the tubes can be controlled independently over large ranges. Such soft composite materials offer a potential interest for the synthesis of mesoporous materials.

MANUSCRIPT TEXT

Surfactant- and polymer-assisted templating strategies are very successful approaches[1-3] for the synthesis of ordered mesoporous materials, which are potentially attractive for applications in catalysis, separation, optics, and photonics. The mesoporous materials display often a hexagonal symmetry with one-dimensional pores arranged on a triangular lattice in the oxide medium. In true liquid crystal templating routes[4], syntheses are directly performed within a pre-formed liquid crystal template. Because one can control the symmetry and the characteristic sizes of the soft template (radius of the tubes and spacing between the tubes, for a hexagonal phase for instance), this approach provides in principle a more predictable route than syntheses in dilute micellar solutions. One expects indeed the symmetry and the characteristic sizes of the as-synthesized mesoporous material to reflect those of the template. Very often however, binary mixtures are used for soft templates. This considerably restricts the range of pore size and of thickness of the walls between adjacent pores available, for the mesoporous materials synthesized in these templates. Attempts to overcome these limitations have been carried out by mixing two types of copolymer or a copolymer and a cosurfactant[5-6] but without clear success so far.

We investigate the effect of amphiphilic copolymers on the stability of hexagonal microemulsions and on the range of lattice parameter accessible. The system consists in surfactant-stabilized oil tubes, of tunable radius, $R$, arranged on a hexagonal lattice in water. We have previously shown that we can control $R$ over one decade (from 3 to 30 nm) by tuning concomitantly the spontaneous curvature of the mixed surfactant/co-surfactant monolayer and the oil content[7-8]. The oil-swollen hexagonal phases can be used for the synthesis of mesostructured materials[9-11]. We show in this Letter that the addition of amphiphilic copolymer, that decorates the oil/water interface, can dramatically increase the accessible characteristic length of the hexagonal lattice, by stabilizing soft materials with a thickness of the water channel between adjacent tubes that can be significantly increased (Scheme 1). These novel results are quantitatively interpreted in terms of a copolymer-induced increase of the effective diameter of the tube. This allows an evaluation of the thickness of the polymer layer that decorates the surfactant-stabilized tubes, which is in reasonable numerical agreement with simple arguments.

We use Cetyl Pyridinium Chloride (CpCl) as surfactant, pentanol as co-surfactant, salted water with a NaCl concentration of 0.3 M, and cyclohexane, as oil. CpCl is received from Fluka and is purified by successive recrystallizations in water and acetone. The other compounds are used as received. The diblock copolymer, $C_{18}$-$PEO_{5K}$, is a home-synthesized amphiphilic polymer, constituted of a polyethylene oxide (PEO) central block, of molecular weight 5 000 g/mol, grafted at one extremity to a $C_{18}H_{37}$ aliphatic chain. The PEO has been hydrophobically modified and purified in the laboratory using methods described in the literature[12-13]. The molecular weight of the starting PEO is determined by size-exclusion chromatography. The hydrophobically modified PEO contains an uretane group between the alkyl chain and the ethylene oxide chain. The semi-developed formula is $[CH_3\text{-}(CH_2)_{17}]\text{-}NH\text{-}CO\text{-}(CH_2CH_2O)_{113}\text{-}O\text{-}CH_3$. After modification, the degrees of substitution of the hydroxyl groups are determined by NMR[14] and are found to be larger than 98%. The radius of gyration of the POE block is $R_G$= 2.4 nm[15]. Additional tests have been performed with two other polymers: a commercial Symperonics F108 by Serva and a triblock copolymer, $C_{18}$-$PEO_{10K}$-$C_{18}$. F108 is an amphiphilic copolymer of formula $PEO_{127}$-$PPO_{48}$-$PEO_{127}$, where EO is ethylene oxide and PO is propylene oxide, and $C_{18}$-$PEO_{10K}$-$C_{18}$ is a home-synthesized telechelic polymer, constituted of a polyethylene oxide (PEO) central block, of molecular weight 10 000 g/mol, grafted at each extremity to a $C_{18}H_{37}$ aliphatic chain. We have checked that a stabilizing effect is also observed with these two polymers. Note that we use a cationic surfactant and not an anionic one (the oil-swollen hexagonal phases are equally stable with Sodium Dodecyl Sulfate -SDS- as surfactant) in order to avoid attractive interactions between the water soluble chains and the surfactant heads[16-17]. We have indeed checked that, although the phase-diagram of the SDS-based hexagonal phases is analogous to the one for CpCl-based hexagonal phases[8], SDS-based hexagonal phases are not stable upon addition of copolymer. We believe that this is due to the attractive interactions between the POE chain of the polymer and the SDS polar heads. Indeed, it has been recognized for a long time that the interactions between cationic surfactants and uncharged water-soluble polymers are very weak, if not totally absent, whereas attractive interactions between the same polymers and anionic surfactants do exist[16,17]. Attractive interactions lead to adsorption of PEO chains that can modify both the bending moduli and the spontaneous curvature of the surfactants films and consequently the morphologies of the aggregates[18].

In all experiments, the weight ratio of oil over surfactant, *O/S*, is fixed, in order to ensure that *R* is constant. We take *O/S*=3.89. We quantify the swelling of the hexagonal phase through the weight ratio of water over surfactant, *W/S*. We define $\beta$ as the molar ratio of polymer over surfactant. In these experiments, $\beta$ varies between 0 and 3%. The parameter $\beta$ is directly related to the surface coverage of the tubes by the copolymer. The cross-over between the mushroom regime and the brush regime[19] is evaluated at $\beta$*=2.2%.

To prepare swollen hexagonal phases, the surfactant and copolymer are first dissolved in the aqueous salted medium, giving a transparent and viscous micellar solution. The addition under stirring of the oil into the pristine micellar solution leads to a white unstable emulsion. The co-surfactant is then added to the mixture, which is vortexed for a few minutes, and a perfectly transparent gel is obtained. Hexagonal phases (especially the highly swollen ones) are extremely sensitive to the quantity of co-surfactant: for an amount of pentanol smaller than the one required for stabilizing a hexagonal phase, a mixture of emulsion and hexagonal phase is obtained; a slightly larger amount than the one required for stabilizing a hexagonal phase leads to the occurrence of a lamellar phase. We show in Figure 1 a plot of the molar ratio of co-surfactant (pentanol) over surfactant (CpCl), $n_{cosurf}/n_{surf}$, as a function of the water content, *W/S*, for the hexagonal samples investigated. The linear growth of $n_{cosurf}/n_{surf}$ indicates that, independently of the amount of polymer in the mixtures, the quantity of pentanol required to stabilize a hexagonal phase increases linearly with the quantity of water in the hexagonal phases. This suggests that a fixed quantity of pentanol is located at the oil/water interface (this quantity is equal to the intercept of the linear fit and corresponds to $n_{cosurf}/n_{surf}$=1.17), while the remaining pentanol is dissolved in the water phase, with a constant concentration in water. From the slope of the linear variation of $n_{cosurf}/n_{surf}$ with *W/S* (Fig. 1), one finds that 0.3g of pentanol is dissolved per g of water. This value is almost 10 times larger than the solubility of pentanol in pure water, a discrepancy that could be due to the fact that surfactant molecules are dissolved in the aqueous medium.

The hexagonal phases are perfectly transparent and birefringent gels, whose structure is determined by small-angle X-ray scattering (SAXS) experiments. A typical pattern of a swollen hexagonal phase is shown Figure 2. The scattering pattern consists in diffraction peaks whose positions are in the ratio $1:\sqrt{3}:2:\sqrt{7}$ and correspond respectively to the (10), (11), (20) and (21) diffraction planes. The position of the first peak, $q_0$, allows a direct determination of the center-center distance between adjacent tubes ($d_c$ in Scheme 1) according to $d_c = \frac{2}{\sqrt{3}} \frac{2\pi}{q_0}$. Given the triangular symmetry, the radius of the tubes, *R*, can be calculating from the lattice parameter, $d_c$, and the volume fraction of the non-polar species, $1-\phi_p$:
$R = \left[ \frac{\sqrt{3}}{2\pi}(1-\phi_p) \right]^{1/2}$, where $\phi_p$ is the volume fraction of the polar species. The polar species include the water, the surfactant (CpCl) heads, part of the co-surfactant (pentanol) heads (such that the molar ratio of co-surfactant over surfactant is equal to 1.17), the remaining of the pentanol molecules and the POE block of the diblock polymer. Note that we have chosen to include the surfactant tails in the non-polar medium. Hence the calculated radius *R* includes the hydrophobic tails. The "water thickness", *h*, which comprises the water plus the heads of the surfactant at the oil/water interfaces is simply obtained through: $h = d_c - 2R$.

We find that the maximum swelling (above which a fluid isotropic phase is obtained) increases as the ratio of polymer over surfactant, $\beta$, increases. Without polymer, the maximum swelling corresponds to $W/S=3.5$. Thanks to the incorporation of the polymer, the maximum swelling reaches respectively $W/S=5.3$, 8.6 and 12.5, for $\beta=1$, 2, and 3% respectively. Hence a threefold increase of the capability of the materials to incorporate water is reached for $\beta = 3\%$. Improving even more the swelling capability by increasing the copolymer amount in the mixture was not possible. This demonstrates that the copolymer-induced swelling is effective only for a decorated fluid interface (copolymer in the mushroom regime) and not for a more rigid interface (copolymer in the brush regime). The insert of Figure 2 displays the first order Bragg diffraction peak for several samples with various water and polymer contents. The shift towards smaller $q$ as $W/S$ increases is the signature of an increase of the lattice parameter. In Figure 3, we have plotted $R$ and $h$ as a function of $W/S$, for samples both with and without copolymer: while the radius of the oil tubes remains roughly constant ($R$ is of the order of 10 nm; our calculation shows a decrease of $R$ of about 10%), $h$ is measured to increase linearly with $W/S$. The extrapolation towards 0 yields a thickness of 0.9 nm, in good agreement with the diameter of the CpCl heads. Interestingly, the thickness of the water channel between adjacent tubes can be tuned, independently of the radius of the tubes, over half a decade (from 2.8 nm without polymer up to 14.4 nm for $\beta=3\%$).

The co-surfactant allows a fine tuning of the spontaneous curvature of the surfactant film. As shown previously for samples without copolymer[7], a given co-surfactant over surfactant ratio is required to stabilize a hexagonal phase with large tubes. As discussed above, the linear variation of the ratio of co-surfactant over surfactant with the quantity of water (Figure 1) presumably reflects the partial solubilization of pentanol in water. In addition, the fact that $n_{cosurf}/n_{surf}$ depends only on the amount of water in the mesophase, and does not depend on the copolymer quantity, suggests that the addition of amphiphilic polymer does not modify the spontaneous curvature of the surfactant/co-surfactant film. Instead, we show below that the stabilization effect due to the copolymer can be rationalized by an increase of the effective size of the tubes.

The swelling effect can be quantified by the surface per unit cell occupied by the tubes in a plane perpendicular to the long axis of the tubes. From purely geometrical arguments (see Scheme 1), one can calculate that this surface per unit cell, named thereafter surface density of tubes, reads: $\rho=2\pi 3^{-1/2}(R/d_c)^2$. In Figure 4, the surface density of tubes is plotted as a function of $W/S$. Without copolymer, we find that the minimum density is $\rho_0=0.615$, a value in agreement with theoretical evaluation of the density of tubes in hexagonal phase where excluded volume interaction between the tubes is considered[20-21]. This minimum density is measured to decrease sensibly thanks to the addition of amphiphilic polymers. We found minimum densities of 0.513, 0.374, and 0.312, for $\beta=1\%$, $\beta=2\%$ and $\beta=3\%$. We consider that, upon addition of copolymer, the minimum density of tubes for the hexagonal phase is constant, provided that the radius of the tubes is increased by a polymer layer of thickness, $e$. This allows us to evaluate $e$, following: $\rho_0/\rho_{measur.}=(1+e/R)^2$, where $\rho_0=0.615$ is the experimental value without polymer, and $\rho_{measur.}$ is the measured minimal density for a given copolymer over surfactant ratio. We find that the effective polymer thickness increases continuously with the amount of copolymer, from 1 nm for $\beta=1\%$ to 4.1 nm for $\beta=3\%$ (Insert of Figure 4). To a first order approximation, the radius of the copolymer-decorated tubes is expected to be a weighted average of the radius of the naked tube ($R$) and of the radius of the tubes that comprises a dense shell of copolymer mushrooms ($R+2R_G$). Hence, $e$ is expected, below the overlap ratio $\beta^*$, to vary linearly with the amount of copolymer $\beta$ and to be equal to 2 times $R_G$ at $\beta= \beta^*$. This simple expectation is plotted as a continuous line in the insert of Figure 4 and reproduces reasonably well the experimental data. We note that an analogous increased repulsion has been previously observed for surfactant aggregates of various morphologies decorated by amphiphilic copolymers[22-24].

To conclude, we have shown that small amounts of polymer adsorbed at oil/water interfaces can greatly enhance the stability of swollen surface hexagonal phases, comprising oil tubes in a water matrix. We note that our results present some interesting analogies with the "boosting effect", that is observed when copolymers are added to bicontinuous surfactant microemulsions and that has been interpreted in terms of a variation of the surfactant film curvature elasticity[25-26]. The effect of copolymer on the bending energy of the surfactant films cannot be checked for hexagonal phases. We have proposed a simple and alternative interpretation for the novel copolymer-induced stabilizing effect of highly swollen hexagonal phases, based of an enhanced repulsive interaction between the surfactant tubes decorated by copolymer. We believe that this physical mechanism of purely steric origin is quite general and should be valid for surfactant phases with other topologies: decorating a surfactant aggregate by amphiphilic copolymer should always lead to a stabilizing effect under dilution, provided the copolymer does not modify the morphology of the surfactant aggregates.

In copolymer-stabilized hexagonal mesophases, both the radius of the tubes and the thickness of the water channel between adjacent tubes can be controlled independently over large ranges. The liquid crystalline soft composite materials designed in this Letter are therefore attractive candidates for the synthesis of mesoporous materials using a true liquid crystal templating approach. We are currently investigating this issue.

ACKNOWLEDGMENT: We acknowledge financial support from the European NOE "SoftComp" (NMP3-CT-2004-502235). We thank ESRF for beam time allocation, and T. Phou and R. Aznar for the polymer synthesis.

FIGURES

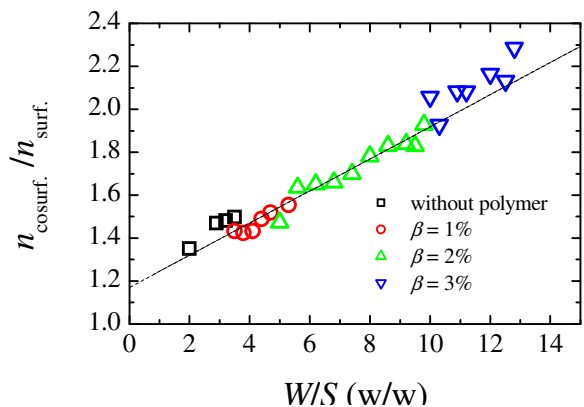

**Figure 1**. Molar ratio of co-surfactant (pentanol) over surfactant (CpCl) for hexagonal phases samples without polymer (black squares) and with a molar ratio of polymer over surfactant (red circles) $\beta = 1\%$, (green up-triangles) $\beta = 2\%$, and (blue down-triangles) $\beta = 3\%$. The line is a linear fit of the data.

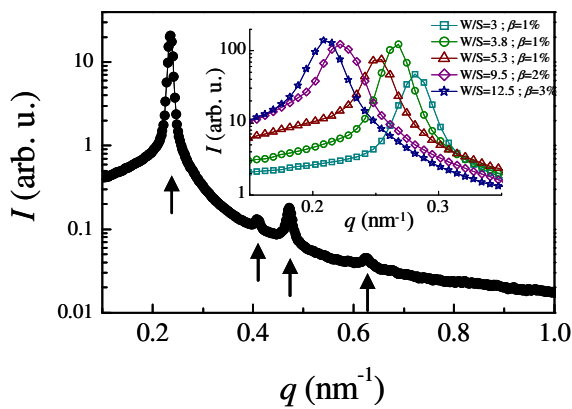

**Figure 2.** SAXS pattern for a sample with water content $W/S = 8$ and copolymer content $\beta = 2\%$. The arrows point the position of the Bragg peaks, which are in the ratio $1:3^{1/2}:2:7^{1/2}$. Insert: Zoom of SAXS patterns around the first order Bragg peak for several samples with various water and polymer contents, as indicated in the caption. The concentrations CpCl/copolymer/pentanol/cyclohexane/water (w/w) for the different samples are 11.9:1.8:4.1:46.4:35.8 for $W/S=3$ and $\beta=1\%$ ; 10.9:1.6:3.8:42.3:41.4 for $W/S=3.8$ and $\beta=1\%$ ; 9.3:1.4:3.6:36.3:49.4 for $W/S=5.3$ and $\beta=1\%$ ; 6.6:1.9:3.1:25.7:62.7 for $W/S=9.5$ and $\beta=2\%$ ; 5.5:2.4:2.8:21.2:68.1 for $W/S=12.5$ and $\beta=3\%$.

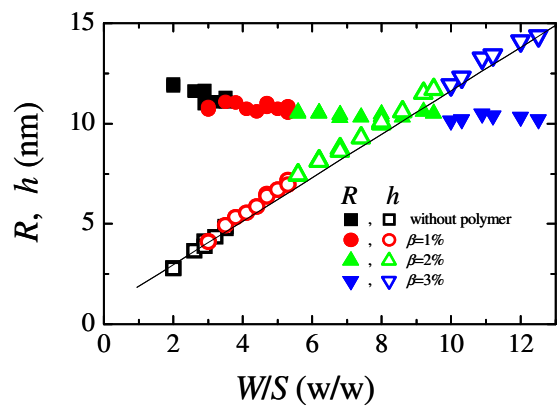

**Figure 3.** Radius of the tubes, $R$, and thickness of the water channel between adjacent tubes, $h$, as a function of the water over surfactant weight ratio for samples without polymer (black squares) and with a molar ratio of polymer over surfactant (red circles) $\beta = 1\%$, (green up-triangles) $\beta = 2\%$, and (blue down-triangles) $\beta = 3\%$. The line is a linear fit of the data.

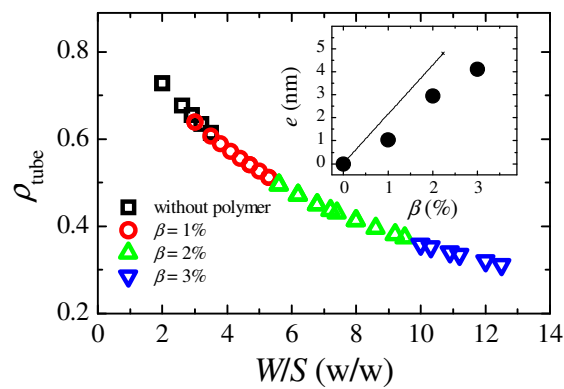

**Figure 4.** Surface density of tubes as a function of the water over surfactant weight ratio for samples without polymer (black squares) and with a molar ratio of polymer over surfactant (red circles) $\beta = 1\%$, (green up-triangles) $\beta = 2\%$, and (blue down-triangles) $\beta = 3\%$. Insert: (symbols) effective thickness of the polymer layer as a function $\beta$, and (line) simple theoretical expectation.

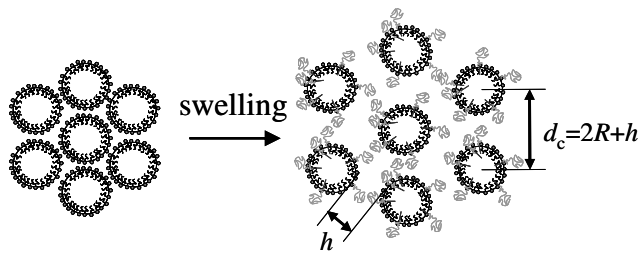

**Scheme 1.** Scheme illustrating the swelling and stabilizing effect of amphiphilic block copolymer on surfactant hexagonal phases.